\documentclass[pra,twocolumn]{revtex4-1}

\usepackage{hyperref}

\begin{document}

\title{The probe readout and quantum limited measurements}
\author{Salini Jose}
\affiliation{School of Physics, Indian Institute of Science Education and Research Thiruvananthapuram, Kerala, India 695016}
\author{Noufal Jaseem}
\affiliation{School of Physics, Indian Institute of Science Education and Research Thiruvananthapuram, Kerala, India 695016}
\author{Anil Shaji}
\affiliation{School of Physics, Indian Institute of Science Education and Research Thiruvananthapuram, Kerala, India 695016}

\begin{abstract}
Assuming that the parameter dependent evolution, as well as the measurements that are done for readout, of a quantum system that acts as the probe in a quantum limited measurement scheme are both fixed, we find the optimal initial states of the probe that will saturate the quantum Cramer-Rao bound. When the probe system is itself made of identical, elementary, two-level subsystems or qubits, we connect the optimal state of the $N$ qubit probe to that of the one qubit probe. This is done for two different classes of dynamics for the probe qubits, one of which is entangling while the other is not. We study the limitations placed on the optimal initial state of the probe and the achievable measurement uncertainty by restrictions on the readout procedure that is applied on the probe qubits at the end of the measurement protocol. 
\end{abstract}

\maketitle

\section{Introduction}

In modeling quantum limited measurements - in particular single parameter estimation - a quantum system that acts as the probe interacts with the measured system in a way that depends on the value of the parameter of interest \cite{giovannetti_quantum-enhanced_2004,giovannetti06a,caves_quantum-circuit_2010,bollinger96a,huelga_improvement_1997}. Knowing the change in the state of the probe that is generated by a parameter dependent Hamiltonian, an estimate of the value of the parameter is obtained. However there is scope to fall into an infinite regression at this step in the modeling of the measurement because, after all, knowing the state of the probe means a measurement of the parameters that describe the state of the probe. So the question, what or who measures the probe, and how is immediately thrown up. Part of the answer lies in the observation that complete knowledge about the state of the quantum probe is not required to obtain the value of the measured parameter. Furthermore the probe is typically assembled by putting together a large number of elementary quantum systems with low dimensional Hilbert spaces. For instance, without loss of generality, one can assume that the probe is made up of $N$ two level quantum systems or qubits~\cite{giovannetti_quantum-enhanced_2004}. The results of vonNeumann type projective measurements~\cite{neumann_mathematical_1996} on to the small, countable, set of basis states of the individual probe units is sufficient to estimate the possibly irrational value of the parameter of interest to any degree of accuracy. The accuracy being dependent on the nature of the quantum probe, the number of elementary units in it and the number of times the probe is applied. To avoid confusion with the measurement of the parameter, we will refer to the measurements on the probe itself as a readout of the probe. 

The readout process is not fully understood either.  This is evidenced by over a century worth of discussions and literature on topics ranging from the collapse of the wave function to decoherence and pointer states~\cite{wheeler_quantum_1983,zurek_decoherence_2003}. However there is not much debate as to whether such readouts can be performed in the laboratory or not, since it indeed is routinely done. Even more sophisticated readouts that include POVMs can be done in the lab and they ultimately boil down to making a larger number of projective measurements on the probe~\cite{neumark47a,peres_neumarks_1990}. 

The motivation for this Paper is the observation that not many types of readouts are possible given the available technology in the design of a quantum limited metrology experiment. For instance, in virtually all precision measurements using interferometers and light, the readout of the state of the light at the output ports is limited to photon counting~\cite{abbott09a,chiruvelli09a,dorner09a,dowling08a,Dowling09a,giovannetti_quantum-enhanced_2004,sanders_optimal_1995}. There are several ways of using photon counting to accomplish readouts of parameters, other than photon number, associated with the state of the light in an interferometer like relative phase etc by combining known transformations of the state with photo counts. Homodyne and heterodyne measurements~\cite{Hariharan:1985pi,Fox:2006fq,Gasvik:2002ef} on light are two examples of such techniques. In particle based metrology protocols like Ramsey interferometry also,  the output state of the spins or atoms are subject to either Stern-Gerlach type or fluorescence type readouts~\cite{Berman:1997rp,gleyzes_quantum_2007}. 

Given the restrictions on the types of readouts that are possible, we find the optimal input state of a quantum probe made of many qubits for a fixed readout procedure and for different choices for the parameter dependent evolution of the probe. In the next section we review the quantum Cramer-Rao bound that forms the basis for the formulation of the problem. In section III the optimal input state for a non-entangling evolution is found. The optimal input state for an entangling evolution is discussed in Section IV followed by a discussion of our results in Section V.  

\section{Saturating the quantum Cramer-Rao Bound \label{sec2}}

The quantum Cramer-Rao bound~\cite{braunstein94a,braunstein96a,helstrom76a,holevo82a} gives the theoretical lower bound on the measurement uncertainty in the estimate of a single parameter $X$ as
\begin{equation}
	\label{eq:CR1}
	\delta X \geq \frac{1}{\sqrt{{\cal F}(X)}} \geq \frac{1}{\sqrt{\langle {\cal L}^{2} \rangle}},
\end{equation}
where ${\cal F}$ is the quantum Fisher information. The symmetric logarithmic derivative operator ${\cal L}$ is defined implicitly by the equation,
\begin{equation}
	\label{eq:symlog1}
	\frac{1}{2} ( {\cal L} \rho_{X} + \rho_{X} {\cal L}) = \frac{d\rho_{X}}{dX} \equiv \rho_{X}'.
\end{equation}
The measurement uncertainty $\delta X$ is quantified using the units corrected, root-mean-squared deviation of the estimate of $X$ from its true value:
\[ \delta X = \frac{X_{\rm est}}{|d\langle X_{\rm est} \rangle_{X}/dX|} - X. \]

Let a general readout on the probe be described by a POVM with a one parameter family of elements $E(\xi)$ such that
\[ \int d\xi E(\xi) = \openone. \]
Let $p(\xi|X)= {\rm tr}(E(\xi) \rho_{X})$ be the measured probabilities for various outcomes of the POVM when the true value of the measured parameter is $X$. As shown in~\cite{braunstein94a}, the quantum Fisher information  is given by
\begin{equation}
	\label{eq:Qfisher}
	{\cal F} = \max_{\{E(\xi) \}} \, F,
\end{equation}
where $F$ is the classical Fisher information computed from the probability distribution for the measurement outcomes as
\begin{eqnarray*}
F & = & \int d\xi \, p(\xi|X) \bigg[\frac{d \ln p(\xi|X)}{d X} \bigg]^{2}  \\
& = &  \int d\xi\, \frac{1}{p(\xi|X)} \bigg[ \frac{d p(\xi|X)}{d X} \bigg]^{2}. 
\end{eqnarray*}
The maximization in the Eq.~(\ref{eq:Qfisher})  is over all possible readout procedures (POVMs) on the probe. Such a maximization is indeed a daunting task and even if it can be done, implementing the POVM that maximizes the Fisher information, thereby minimizing the measurement uncertainty, may, in all likelihood, be impossible to implement in the lab. The second inequality in (\ref{eq:CR1}) circumvents the maximization problem by placing an upper bound on ${\cal F}$ in terms of the expectation value of the square of the symmetric logarithmic derivative operator ${\cal L}$. This expectation value can be computed directly from the initial state of the probe and its parameter dependent dynamics, independent of the readout procedure.  In the context of the current Paper it is worth reprising the sequence of steps detailed in~\cite{braunstein94a} that lead to this upper bound. We have
\begin{eqnarray}
	F & = & \int d\xi \frac{1}{{\rm tr}(E(\xi) \rho_{X})} \bigg[ \frac{d\;}{dX} {\rm tr}(E(\xi) \rho_{X}) \bigg]^{2}, \nonumber \\
	& = & \int d\xi \frac{1}{{\rm tr}(E(\xi) \rho_{X})} \big[ {\rm tr}(E(\xi) \rho_{X}') \big]^{2} ,\nonumber  \\
	& = & \int d\xi \frac{1}{{\rm tr}(E(\xi) \rho_{X})} \bigg[ \frac{1}{2}  {\rm tr}  \big( E(\xi) {\cal L} \rho_{X}  + E(\xi) \rho_{X} {\cal L}\big) \bigg]^{2}, \nonumber \\
	& = &  \int  d\xi \frac{1}{{\rm tr}(E(\xi) \rho_{X})} \bigg[ {\rm Re} \Big(  {\rm tr}  \big(\rho_{X} E(\xi) {\cal L}  \big) \Big) \bigg]^{2},
\end{eqnarray}
where we have used Eq.~(\ref{eq:symlog1}), the cyclic nature of the trace and the Hermiticity of ${\cal L}$ and $E(\xi)$ to obtain the last equality in the equation above. Our focus is on the case where the second inequality in~(\ref{eq:CR1}) is saturated.  When
\begin{equation}
	\label{eq:CR2}
	{\rm Im} \Big[  {\rm tr}  \big(\rho_{X} E(\xi) {\cal L}  \big) \Big ] = 0,
\end{equation}
we have
\begin{eqnarray}
	\label{eq:symmlog2}
	F & = & \int d\xi \frac{1}{{\rm tr}(E(\xi) \rho_{X})} \bigg| {\rm tr}  \big(\rho_{X} E(\xi) {\cal L}  \big) \bigg|^{2} \nonumber \\
	& = &   \int d\xi \bigg| {\rm tr} \bigg( \frac{\sqrt{\rho_{X} E(\xi)}  }{\sqrt{{\rm tr}(E(\xi) \rho_{X}) }} \sqrt{E(\xi)} {\cal L} \sqrt{\rho_{X}} \bigg) \bigg|^{2}. \!\!
\end{eqnarray}
Using the Schwarz inequality, $|\langle x, y\rangle |^{2}\leq \langle x, x \rangle \cdot \langle y, y \rangle$ for the trace norm, we can write an upper bound on the classical Fisher information as, 
\begin{equation}
	\label{eq:symmlog3} 
	F \leq   \int d\xi \, {\rm tr} \bigg( \frac{ E(\xi) \rho_{X}  }{{\rm tr}(E(\xi) \rho_{X}) } \bigg) \cdot {\rm tr} (E(\xi) {\cal L}^{2} \rho_{X}). 
\end{equation}
The Schwarz inequality is saturated when
 \begin{equation}
 	\label{eq:CR4}
	\sqrt{E(\xi) \rho_{X}}= \lambda_{\xi} \sqrt{E(\xi)} {\cal L} \sqrt{\rho_{X}},
\end{equation}
for all $\xi$. Assuming Eq.~(\ref{eq:CR4}) holds, we get
\begin{equation}
	\label{eq:CR5}
	{\cal F} = \int d\xi \,  {\rm tr} (E(\xi) {\cal L}^{2} \rho_{X}) = {\rm tr} ({\cal L}^{2} \rho_{X})  = \langle {\cal L}^{2} \rangle,
\end{equation}
 and the second inequality in~(\ref{eq:CR1}) is saturated. Equations~(\ref{eq:CR2}) and (\ref{eq:CR5}) furnish the conditions on the the readout procedure (POVM) such that a quantum probe in the initial state $\rho_{X}$ that undergoes the parameter dependent evolution implicitly contained in ${\cal L}$ will attain the quantum Cramer-Rao bound. Multiplying Eq.~(\ref{eq:CR4}) by $\sqrt{E(\xi)}$ from the left and $\sqrt{\rho_{X}}$ from the right we obtain,
\begin{equation}
	\label{eq:CR6}
	E(\xi) \bigg( {\cal L} - \frac{1}{\lambda_{\xi}} \openone \bigg) \rho_{X} = 0,
\end{equation}
for any $\rho_{X}$. The above equation is satisfied if the readout is taken to be a set of orthogonal projectors, $E(\xi)$, on to the complete set of orthonormal eigenstates of ${\cal L}$. The $\lambda_{\xi}$ are inverses of the eigenvalues of ${\cal L}$ with 
\begin{equation}
	\label{eq:lambda1} \frac{1}{\lambda_{\xi}} = \frac{ {\rm tr} \big( E(\xi) {\cal L} \rho_{X} \big)}{{\rm tr} \big( E(\xi) \rho_{X}\big)}. 
\end{equation}
Condition~(\ref{eq:CR2}) implies that $\lambda_{\xi}$ are real.

In situations where the readout procedure is fixed due to practical reasons or otherwise, one can now formulate the problem of finding the optimal initial state of the probe $\rho_{X}$ as follows. We limit ourselves to the rather common case where the readout is a complete set of orthogonal projective measurements. Even if the readout is realized by a more general POVM, we assume that a suitable Neumark extension~\cite{neumark47a,peres_neumarks_1990} has been used to reduce it to complete set of orthonormal projectors. Given this complete set of orthogonal projectors denoted as $ \{ |\xi \rangle \}$, we can construct the symmetric logarithmic derivative operator corresponding to this readout procedure as
\[ {\cal L} = \sum_{\xi} \frac{1}{\lambda_{\xi}} |\xi \rangle \langle \xi|. \]
The optimal initial state of the probe $\rho_{X}$ for this readout satisfies the equation
\begin{equation}
	\label{eq:sol1}
	\frac{1}{2} ({\cal L} \rho_{X} + \rho_{X} {\cal L}) = \rho' =- i[H, \, \rho_{X}],
\end{equation}
assuming that the parameter dependent evolution of the probe is generated by the Hamiltonian,
\begin{equation}
	\label{eq:Hamil}
	H_{\rm probe} = X H.
\end{equation}

In the following we will investigate the solutions of Eq.~(\ref{eq:sol1}) for two choices of $H$, one that entangles the probe qubits and one that does not. The number of probe qubits $N$ is the resource against which the performance of the measurement scheme is calibrated. The discussion here is quite general and can be applied to the case where the number of probe units itself is not the most important resource. For instance, in interferometry with light the circulating power in the interferometer and not the number of photons is the crucial, limited resource~\cite{abbott09a}. In other words, in our discussion, $N$ is essentially a place holder for the relevant resource for each measurement scheme and a mapping between the real resource and $N$ can be found quite easily for most quantum limited metrology schemes. 

\section{Non-entangling evolution of the probe qubits \label{sec3}}

Let the parameter independent part of the Hamiltonian in~(\ref{eq:Hamil}) that governs the time evolution of the $N$ qubit probe have the form
\begin{equation}
	\label{eq:Hamil2}
	H = \sum_{j=1}^{N} h^{(j)}, 
\end{equation}
where $h^{(j)}$ is an operator that acts only on the $j^{\rm th}$ qubit. By construction the time evolution generated by this Hamiltonian will not lead to entanglement between the probe qubits. Without loss of generality, using the freedom to define a basis independently for the two dimensional Hilbert spaces of each of the individual qubits, we choose all the single qubit operators to be identical and equal to 
\[ h^{(j)} = \frac{1}{2} \sigma_{3}^{(j)}. \]

Once we choose to define the basis for the Hilbert space of each qubit so that the evolution Hamiltonian is as given above, we make the assumption that the read out procedure is limited due to practical considerations or otherwise to projective measurements along the states $|+\rangle$ and $|-\rangle$ for each qubit where
\begin{equation}
	\label{eq:readout1} 
	| \pm \rangle = \frac{1}{\sqrt{2}} (|0\rangle \pm | 1\rangle). 
\end{equation}
The symmetric logarithmic derivative operator for which this readout procedure saturates the quantum Cramer-Rao bound is then
\[ {\cal L} = \sum_{j_{1},j_{2} \ldots j_{N}} \frac{1}{\lambda_{j_{1}j_{2} \ldots j_{N}}} |j_{1} j_{2} \ldots j_{N} \rangle \langle j_{1}j_{2} \ldots j_{N} |, \]
where $j_{l} = \{+, -\}$. The optimal state of the $N$ qubit probe corresponding to this dynamics and readout can now be found by solving Eq.~(\ref{eq:sol1}).

We first look at the case where the probe is made of a single qubit; i.e.~$N=1$. In this case
\[ {\cal L} = \frac{1}{\lambda_{+}}|+\rangle \langle + | + \frac{1}{\lambda_{-}} |-\rangle \langle -|. \]
We write an arbitrary state of the probe as
\[ \rho_{X} = \frac{1}{2}(\openone + a_{i} \sigma_{i}), \]
where $\sigma_{i}$, $i=1,2,3$ are the Pauli matrices and the dependence of the state on $X$ is hidden in the dependence of the coefficients $a_{i}$ on the estimated parameter. Using
\[ |+ \rangle \langle + | = \frac{1}{2}(\openone + \sigma_{1}) \quad {\rm and} \quad |-\rangle \langle -| = \frac{1}{2}(\openone - \sigma_{1}), \]
and the anti commutation relations of the Pauli matrices we have
\begin{eqnarray}
	\label{eq:lhs1}
	\frac{1}{2}\{{\cal L}, \rho_{X} \} & = & \frac{1}{4} \bigg[ \bigg( \frac{1+a_{1}}{\lambda_{+}} + \frac{1-a_{1}}{\lambda_{-}}\bigg) \openone \nonumber \\
	&&  \quad + \;  \bigg( \frac{1+a_{1}}{\lambda_{+}} - \frac{1-a_{1}}{\lambda_{-}}\bigg) \sigma_{1} \nonumber \\
	&&  \quad + \; \bigg( \frac{1}{\lambda_{+}} + \frac{1}{\lambda_{+}} \bigg) a_{2} \sigma_{2} \nonumber \\
	&&  \quad + \; \bigg( \frac{1}{\lambda_{+}} + \frac{1}{\lambda_{+}} \bigg) a_{3} \sigma_{3}\bigg], 
\end{eqnarray}
and
\begin{equation}
	\label{eq:rhs1}
	-i[H, \, \rho_{X}] = -\frac{1}{2} (a_{2} \sigma_{1} - a_{1} \sigma_{2}). 
\end{equation}
Inserting Eqs.~(\ref{eq:lhs1}) and (\ref{eq:rhs1}) into Eq.~(\ref{eq:symlog1}) we find a solution as
\begin{equation}
	\label{eq:lambda2}
	\frac{1}{\lambda_{+}}= -\frac{1}{\lambda_{-}} = - a_{2}, \qquad {\rm with} \qquad a_{1} =0.
\end{equation}
Since $a_{3}$ does not appear in the right side of the equation, we are free to choose its value depending on the available state preparation procedure for the probe qubit.  Using Eq.~(\ref{eq:lambda2}) we have
\[ {\cal L} =- a_{2}|+\rangle \langle +| + a_{2}|-\rangle \langle -| = - a_{2} \sigma_{1}. \]
The quantum Fisher information is therefore
\[ {\cal F} = \langle {\cal L}^{2} \rangle = {\rm tr} (a_{2}^{2} \openone \rho) =  a_{2}^{2}. \] 
The choice, $a_{2} = \pm 1$ maximizes ${\cal F}$ and positivity of the state of the probe qubit now requires that $a_{3} = 0$. So we find two possible optimal states of the one qubit probe for the given dynamics and readout as
\begin{equation}
	\label{eq:opt1} 
	\tilde{\rho}_{X} = \frac{1}{2} ( \openone \pm \sigma_{2}). 
\end{equation}

The results obtained so far are not anything new or unexpected. In Ramsey interferometers~\cite{gleyzes_quantum_2007, bdfsbc08} using atoms with two effective states in play forming qubits, the effective evolution of the probe units is modeled as rotations about the $\sigma_{3}$ axis in the Bloch sphere while the qubits themselves are initialized along the $\sigma_{1}$ or $\sigma_{2}$ directions. The optimal readout is then measurements on the individual probe qubits along $\sigma_{2}$ or $\sigma_{1}$ directions respectively. The point of the preceding discussion is primarily to illustrate the means of obtaining the optimal state of the probe given that the readout is fixed. 

For dynamics generated by a non-entangling Hamiltonian of the form given in Eq.~(\ref{eq:Hamil2}), it is known that the Heisenberg limited scaling of $1/N$ is obtained for a ``Schr\"{o}dinger cat'' state that is highly entangled~\cite{bollinger96a,caves_quantum-circuit_2010,huelga_improvement_1997}. However this assumes the ability to do a phase kick back operation after the parameter dependent evolution of the probe followed by projective measurements in order to implement, in effect, a readout on to a basis of entangled states.  If we restrict the readout on each probe qubit to be along the basis given in~(\ref{eq:readout1}), then we have to again use the approach discussed above to find the optimal initial state of a multi-qubit quantum probe. 

We briefly discuss the $N=2$ case first. We have
\begin{eqnarray*}
{\cal L}& =& \frac{1}{\lambda_{++}}|++\rangle \langle ++ | + \frac{1}{\lambda_{+-}} |+-\rangle \langle +-|\nonumber \\
   & & + \frac{1}{\lambda_{-+}}|-+\rangle \langle -+ | + \frac{1}{\lambda_{--}} |--\rangle \langle --|. \nonumber \\
\end{eqnarray*}
and
\[ H^{(2)}= \frac{1}{2} (\sigma_3\otimes\openone+\openone\otimes\sigma_3). \]
Using the commutators and anti-commutators for tensor products of Pauli operators given in~\cite{altafini_commuting_2005}, and assuming that the initial state of the two qubits, $\rho^{(2)}_{X}$ has the generic form
\[ \rho^{(2)}_{X} = \frac{1}{4} (\openone \otimes \openone + a_{i} \sigma_{i} \otimes \openone + b_{j} \openone \otimes \sigma_{j} + c_{ij} \sigma_{i} \otimes \sigma_{j}), \]
we obtain sixteen algebraic equations (see Appendix \ref{app1} for details) from Eq.~(\ref{eq:sol1}) by equating coefficients of corresponding operators. It is worth noting that the Schr\"{o}dinger cat states, 
\[ |\Phi_{\pm} \rangle = \frac{1}{\sqrt{2}} (|00\rangle \pm |11\rangle), \]
with the corresponding density matrices
\[ \rho^{(2)}_{\pm} = \frac{1}{4} ( \openone \otimes \openone + \sigma_{1} \otimes \sigma_{1} \pm \sigma_{2} \otimes \sigma_{2} + \sigma_{3} \otimes \sigma_{3}), \]
are not solutions of the equations we obtain. This again is symptomatic of the restriction on the readout procedure we have imposed. On the other hand it is straightforward to verify, as is done in Appendix~\ref{app1} that 
\[ \tilde{\rho}^{(2)}_{X} = \tilde{\rho}_{X} \otimes \tilde{\rho}_{X}, \]
is a solution, where $\tilde{\rho}_{X}$ is the optimal state of the single qubit probe obtained in Eq~(\ref{eq:opt1}).  We find that
\[ {\cal L} = -2|++\rangle \langle ++| + 2 |--\rangle \langle --| = -\sigma_{1} \otimes \openone - \openone \otimes \sigma_{1}, \]
so that
\[ {\cal F} = \langle {\cal L}^{2} \rangle = 2 \langle \openone \otimes \openone + \sigma_{1} \otimes \sigma_{1} \rangle  = 2  = 4 \langle \Delta^{2} H^{(2)} \rangle.\]

Generalizing to $N$ qubits we again find that,
\[ \tilde{\rho}^{(N)}_{X} = \tilde{\rho}_{X}^{\otimes N}, \]
is a solution of Eq.~(\ref{eq:sol1}) with
\[ \frac{1}{\lambda_{j_{1} j_{2} \cdots j_{N}}} = \frac{1}{\lambda_{j_{1}}} + \frac{1}{\lambda_{j_{2}}} + \cdots + \frac{1}{\lambda_{j_{N}}}, \quad j_{l} = \{+, -\}. \]
Detailed proofs of these results are given in Appendices~\ref{app1a} and \ref{app1b}. Significantly, we see that $\langle {\cal L}^{2} \rangle$ scales as $N$ rather than as $N^{2}$. So one does not reach the Heisenberg limited scaling of $1/N$ for the linear, non-entangling, parameter dependent dynamics of the quantum probe. 

The main point of the preceding discussion on a particular example of non-entangling dynamics is to highlight the fact that with a restricted readout procedure, it might not be possible to go beyond the shot noise limited scaling of $1/\sqrt{N}$ for the measurement uncertainty even if the ability to initialize the quantum probe in arbitrary entangled quantum states is available.  In other words, for implementing quantum limited measurements that beat the shot noise limit, we see that devising ways of doing possibly complicated readouts can be as important as control over the initial state of the quantum probe and its dynamics.  

\section{Entangling dynamics \label{entangling}}

Now let us consider entangling dynamics for the probe qubits generated by 
\begin{equation}
	\label{eq:hamil2}
	H^{(N)} = \frac{1}{2} \sigma_{3}^{\otimes N}. 
\end{equation}
Note that this Hamiltonian does not belong to the family of non-linear Hamiltonians discussed in~\cite{boixo_generalized_2007,boixo_quantum-limited_2007} that leads to measurement schemes in which the  uncertainty scales as $1/N^{k}$ or $1/N^{k-1/2}$ with respect to $N$ depending on whether the initial state of the probe is entangled or not. 

The readout procedure in this case is also the same as before with independent measurements of the individual qubits along the $|\pm\rangle$ axis. However, despite this restriction $\tilde{\rho}_{X}^{\otimes N}$ is not, in general a solution to Eq.~(\ref{eq:sol1}). In fact for the entangling Hamiltonian in Eq.~(\ref{eq:hamil2}) one can show that (see Appendix \ref{app2}) if $\tilde{\rho}_{X}^{(2d)}$ for $d=1,2,\ldots$ is not a solution of the same equation giving the optimal state of a quantum probe made of $2d$ (even number of) qubits. 

For $N=2$, it is still worthwhile to find the optimal state even if it cannot be $\tilde{\rho}_{X} \otimes \tilde{\rho}_{X}$. One possible solution for Eq.~(\ref{eq:sol1}) is the state
\begin{equation}
	\label{eq:2qubit1}
	\tilde{\rho}_{X}^{(2)} = \frac{1}{4} (\openone \otimes \openone + c_{11} \sigma_{1} \otimes \sigma_{1} +  c_{23} \sigma_{2} \otimes \sigma_{3} + c_{32} \sigma_{3} \otimes \sigma_{2}).
\end{equation}
This state is pure when  $c_{11} = c_{23} = c_{32} =1 $. However for this state $\langle {\cal L }^{2} \rangle = 1$, indicating that it does not saturate the quantum Cramer-Rao bound for the given dynamics and readout. In fact the restriction that the qubits are measured independently means that even with two qubits, the measurement uncertainty is not improved compared to the single qubit probe. The solution to Eq.~(\ref{eq:sol1}) obtained in (\ref{eq:2qubit1}) is not unique either. For instance, another solution is obtained immediately from the state above by changing the signs of $c_{23}$ and $c_{32}$ which in turn swaps $1/\lambda_{++}$ and $1/\lambda_{--}$. For larger $N$ one can solve the system of algebraic equations generated from Eq.~(\ref{eq:sol1}) by equating the coefficients of corresponding operators to find one or more optimal initial states of the probe that saturate the quantum Cramer Rao bound for the entangling dynamics.

Rather than solving for the optimal state on a case-by-case basis, it is more useful to pursue solutions of the form $\tilde{\rho}_{X}^{\otimes N}$ even though we have ruled out such solutions for even $N$. For odd $N$ we can show that  $\tilde{\rho}_{X}^{\otimes N}$ is a solution of Eq.~(\ref{eq:sol1}) with the $\lambda$'s given by
\[ \frac{1}{\lambda_{j_{1}j_{2} \ldots j_{N}}} = \frac{i^{N+3}}{\lambda_{j_{i}} \lambda_{j_{2}} \cdots \lambda_{j_{N}}}, \]
 where $1/\lambda_{j_{l}}$ are the eigenvalues of ${\cal L}$ for the $N=1$ case. The proof of this result is rather technical and long and is given in Appendix~\ref{app3}. This is not a very useful result because $1/\lambda_{j_{l}}$ have values $\pm 1$ and so $1/\lambda_{j_{1},j_{2} \ldots j_{N}} = \pm 1$ for odd $N$. A simple computation shows that then ${\cal L}^{2} = \openone$ and so $\langle {\cal L}^{2} \rangle = 1$. In other words, no advantage is obtained in having $N$ qubits in the probe rather than one if the $N$ are initialized in the state $\tilde{\rho}_{X}^{\otimes N}$. Even with the entangling evolution, in a real experiment, if the ability to initialize and readout the probe in states that are not simple tensor products is not available, then the reduction in measurement uncertainty promised by quantum limited metrology is wiped out.   
\section{Conclusion}

A general quantum limited measurement for estimating a single parameter can be thought of having three stages. There is a preparation stage in which the quantum system that is acting as the probe of the measured parameter is initialized in a particular quantum state. The second stage is the parameter dependent evolution of the quantum probe and the last stage is the readout of the probe. The advantages of using specific - often entangled - initial states of the quantum probe was explored extensively during the initial phase of the development of the theory and implementation of quantum limited measurement schemes~\cite{bollinger96a,caves80b,caves81a, huelga_improvement_1997,yurke86a,yurke86b}. How the dynamics influences the measurement uncertainty was explored more recently~\cite{boixo_generalized_2007,boixo_quantum-limited_2007,boixo08c,caves_quantum-circuit_2010}.

This paper is focused on the third stage of a quantum metrology scheme when considerations, practical or otherwise, limit the types of readout that can be done on the quantum probe. This analysis is done in the limited context of a qubit based metrology schemes. Extensions to other quantum limited measurement schemes including interferometry with squeezed states, N00N states etc may also be considered. We see that arbitrary state preparations and dynamics might not be particularly useful in delivering an improved measurement uncertainty if there are limitations on the readout. For instance, in the case of the $N$ qubit probe evolving under a non-entangling Hamiltonian, the Schr\"{o}dinger cat state turns out not to be the optimal state because of the restriction that the readout is limited to independent measurements on each of the $N$ qubits. In the case of the entangling dynamics we see that when the initial state of the probe is a product state then with the same restriction as before on the readout, the performance of the measurement scheme is no better than what can be done with a single qubit probe.

\acknowledgments
This work is supported in part by a grant from the Fast-Track Scheme for Young Scientists (SERC Sl.~No.~2786), and the Ramanujan Fellowship programme (No.~SR/S2/RJN-01/2009), both of the Department of Science and Technology, Government of India. Anil Shaji thanks Animesh Datta for preliminary discussions that led to this work. 

\appendix
\section{Non-entangling Hamiltonian \label{app1}}

For two qubits, Eq.~(\ref{eq:sol1}) leads to sixteen equations connecting $\lambda_{++}$, $\lambda_{+-}$, $\lambda_{-+}$ and $\lambda_{--}$ to the fifteen coefficients $a_{i}$, $b_{j}$ and $c_{ij}$ defining the state of the probe. These are obtained by equating the coefficients of operators of the form $\sigma_{\alpha} \otimes \sigma_{\beta}$, $\alpha, \beta = 0, \dots, 4$ with $\sigma_{0} \equiv \openone$.  Using the notation,
\[ \kappa_{\pm \pm \pm}=\bigg( \frac{1}{\lambda_{++}} \pm \frac{1}{\lambda_{+-}}\pm \frac{1}{\lambda_{-+}}\pm \frac{1}{\lambda_{--}}\bigg),\]
the equations we get are
\begin{eqnarray}
	\label{eq:2Qnent}
	K_{+++} + K_{+--} a_{1} + K_{-+-} b_{1} + K_{--+}c_{11}\! & \! =\! &  0 \nonumber \\
	K_{-+-} +K_{--+} a_{1} + K_{+++} b_{1} + K_{+--} c_{11} \! & \!= \!& -4b_{2} \nonumber \\
	K_{+--} + K_{+++} a_{1} + K_{--+} b_{1} + K_{-+-} c_{11} \! & \!=\! & -4a_{2} \nonumber \\
	K_{--+} + K_{-+-}a_{1} + K_{+--} b_{1} + K_{+++} c_{11} \! & \!= \!& - 4c_{12} - 4c_{21} \nonumber \\
	K_{+++} b_{2} + K_{+--} c_{12} & = & 4 b_{1} \nonumber \\
	K_{+--} b_{2} + K_{+++}c_{12} & = &  4c_{11} - 4c_{22} \nonumber \\
	K_{+++} a_{2} + K_{-+-} c_{21} & = & 4 a_{1} \nonumber \\
	K_{-+-} a_{2} + K_{+++} c_{21} & = & 4 c_{11} - 4c_{22} \nonumber \\
	K_{+++} a_{3} + K_{-+-} c_{31} & = & 0 \nonumber \\
	K_{-+-} a_{3} + K_{+++} c_{31} & = & -4c_{32} \nonumber \\
	K_{+++} b_{3} + K_{+--} c_{13} & = & 0 \nonumber \\
	K_{+--} b_{3} + K_{+++} c_{13} & = & - 4c_{23} \nonumber \\
	K_{+++} c_{22} - K_{--+} c_{33} & = & 4 c_{12} + 4 c_{21} \nonumber \\
	K_{+++} c_{23} + K_{--+} c_{32} & = & 4 c_{13} \nonumber \\
	K_{--+} c_{23} + K_{+++} c_{32} & = & 4 c_{31} \nonumber \\
	K_{+++} c_{33} - K_{--+}c_{22} & = & 0.
\end{eqnarray}

For the particular case in which 
\begin{eqnarray*} 
\rho^{(2)}_{X} & = & \tilde{\rho}_{X} \otimes \tilde{\rho}_{X}  \\
& = & \frac{1}{4}(\openone \otimes \openone +  \sigma_{2} \otimes \openone +  \openone \otimes \sigma_{2} + \sigma_{2} \otimes \sigma_{2} ), 
\end{eqnarray*}
the non-trivial equations amongst Eqs.~(\ref{eq:2Qnent}) are
\begin{eqnarray*}
	K_{+++} & = K_{--+} & = 0, \\
	K_{-+-} & = K_{+--} & = -4.
\end{eqnarray*}
From these equations we get
\[ \frac{1}{\lambda_{++}} = -\frac{1}{\lambda_{--}} = -2, \]
and
\[ \frac{1}{\lambda_{+-}} = -\frac{1}{\lambda_{-+}} = 0, \]
as a possible solution. Note that in this case,
\[ \frac{1}{\lambda_{jk}} = \frac{1}{\lambda_{j}} + \frac{1}{\lambda_{k}}, \qquad j, \, k = \{+, \, -\}. \] 

\subsection{N-qubits \label{app1a}}

In \cite{braunstein94a}, an alternate expression for the symmetric logarithmic derivative operator is obtained as
\[ {\cal L}(O) = \sum_{p_{j} + p_{k} \neq 0} \frac{2}{p_{j} + p_{k}}  O_{jk} |j\rangle \langle k|; \qquad O_{jk} = \langle j | O | k \rangle , \]
where $O$ is the operator on which ${\cal L}$ acts. We write the optimal single qubit state as 
\[ \tilde{\rho}_{X}=\sum_j{p_j |j\rangle\langle j|}.\]
In our particular example the basis in which $\tilde{\rho}_{X}$  is diagonal is given by the vectors, $| i\rangle = (|0\rangle + i |1\rangle)/\sqrt{2}$ and  $|\bar{i}\rangle = (|0\rangle - i |1\rangle)/\sqrt{2}$.  In the remainder we want to use the result ${\cal F} = \langle {\cal L}^{2}(\rho') \rangle$, also obtained in \cite{braunstein94a}. For the single qubit probe with evolution generated by $H=\sigma_{3}/2$ we have
\[ \rho'_{jk} = -i \langle j | [H,\, \tilde{\rho}_{X}] |k \rangle = -\frac{i}{2} (p_{k} - p_{j}) \langle j | \sigma_{3} | k \rangle, \]
and
\[ {\cal L} (\rho') = -i\sum_{jk} \frac{p_{k} - p_{j}}{p_{j} + p_{k}} \langle j | \sigma_{3} | k \rangle |j\rangle \langle k|, \]
with the sum extending over all $j$, $k$ such that $p_{j} + p_{k} \neq 0$. If the readout procedure corresponds to projective measurements with elements,
\[ E_{n} = |\theta_{n} \rangle \langle \theta_{n}|, \]
we have
\[ {\rm tr}[\rho E_{n} {\cal L}(\rho')] = -i \sum_{jk}  p_{k} \frac{p_{k} - p_{j}}{p_{k} + p_{j}} \langle j | \sigma_{3} | k \rangle  \langle k | \theta_{n} \rangle \langle \theta_{n}|j\rangle,\]
and
\[ {\rm tr}[\rho E_{n}] = \sum_{j} p_{j} |\langle j | \theta_{n}\rangle|^{2} .\]
This gives us
\[ \frac{1}{\lambda_{n}} = -i \frac{\sum_{jk}  p_{k} \frac{p_{k} - p_{j}}{p_{k} + p_{j}} \langle j | \sigma_{3} | k \rangle  \langle k | \theta_{n} \rangle \langle \theta_{n}|j\rangle}{\sum_{l} p_{l} |\langle l | \theta_{n}\rangle|^{2} }. \]

The $N$ qubit state tensor product state can be written as 
\[ \rho^{(N)}_{X} = \sum_{j_{1},j_{2} \ldots j_{N}} p_{j_{1}} p_{j_{2}} \cdots p_{j_{N}} |j_{1} j_{2} \cdots j_{N} \rangle \langle j_{1} j_{2} \cdots j_{N}|, \]
while the Hamiltonian corresponding to non-entangling evolution on the $N$ qubits is,  
\[ H=\frac{1}{2} (\sigma_3\otimes \openone^{\otimes (N-1)}+\cdots +\openone^{\otimes (N-1)}\otimes\sigma_3). \]
 From $\rho' = - i[H, \, \rho]$, and denoting the string $j_{1}j_{2} \ldots j_{N}$ as $\vec{j}$, we have 
\begin{eqnarray*}
\rho_{\vec{j}\vec{k}}'&=& -i \langle \vec{j}|[H, \, \rho]|\vec{k}\rangle\nonumber\\
&=& -\frac{i}{2} \sum_{l=1}^{N} p_{j_{1}}p_{j_{2}} \ldots p_{k_{l}} \ldots p_{j_{N}} \langle j_{l} | \sigma_{3} | k_{l} \rangle \prod_{n \neq l} \delta_{j_{n} k_{n}}\nonumber \\
&& \quad + \frac{i}{2}\sum_{l=1}^{N} p_{j_{1}}p_{j_{2}} \ldots p_{j_{l}} \ldots p_{j_{N}} \langle j_{l} | \sigma_{3} | k_{l} \rangle \prod_{n \neq l} \delta_{j_{n} k_{n}} \nonumber \\
&=& -\frac{i}{2}\sum_{l=1}^{N} (p_{k_{l}} - p_{j_{l}})  \langle j_{l}|\sigma_3 |k_{l}\rangle  \prod_{m \neq l} p_{j_{m}}  \delta_{j_{m} k_{m}}
\end{eqnarray*}
using the above, we obtain 
\begin{eqnarray*}
{\cal L}(\rho')&=&-i \sum_{l=1}^{N} \sum_{j_{1}, \ldots, j_{l}, \ldots j_{N}, k_{l}} \frac{p_{k_{l}}-p_{j_{l}}}{p_{j_{l}}+p_{k_{l}}}  \langle j_{l}|\sigma_3 |k_{l}\rangle  \\ 
&&\qquad \times \; |j_{1}, \ldots j_{l}, \ldots j_{N} \rangle \langle j_{1}, \ldots k_{l}, \ldots j_{N}|.
\end{eqnarray*}
Assuming that the readout procedure consists of projective measurements corresponding to the operators, 
\[ E_{\vec{n}} = |\theta_{n_{1}} \theta_{n_{2}} \cdots \theta_{n_{N}} \rangle \langle \theta_{n_{1}} \theta_{n_{2}} \cdots \theta_{n_{N}}|, \]
we have
\begin{eqnarray*}
{\rm tr} [\rho E_{\vec{n}} {\cal L }(\rho')]&=&-i \sum_{l=1}^{N} \sum_{j_{1}, \ldots, j_{l}, \ldots j_{N}, k_{l}}  p_{j_{l}}\frac{p_{j_{l}}-p_{k_{l}}}{p_{j_{l}}+p_{k_{l}}}  \langle j_{l}|\sigma_3 |k_{l}\rangle \\
& & \quad \times \; \langle j_{l}|\theta_{n_{l}}\rangle \langle \theta_{n_{l}}|k_{l}\rangle \prod_{m\neq l} p_{j_{m}} |\langle j_{m}|\theta_{n_{m}} \rangle |^2 \\
& = & -i \sum_{l=1}^{N}  \bigg( \sum_{\{j_{m}\}_{m\neq l}} \prod_{m\neq l} p_{j_{m}} |\langle j_{m}|\theta_{n_{m}} \rangle |^2 \bigg)\\
&& \times \bigg( \sum_{j_{l} k _{l}}p_{j_{l}}\frac{p_{j_{l}}-p_{k_{l}}}{p_{j_{l}}+p_{k_{l}}}  \langle j_{l}|\sigma_3 |k_{l}\rangle  \\
&& \qquad \qquad \langle j_{l}|\theta_{n_{l}}\rangle \langle \theta_{n_{l}}|k_{l}\rangle \bigg),
\end{eqnarray*}
where $\{j_{m}\}_{m\neq l}$ stands for $j_{1} \ldots, j_{l-1}, j_{l+1}, \ldots j_{N}$. Similarly we have
\begin{eqnarray*}
{\rm tr}(E_{\vec{n}}\rho) & = & \sum_{j_{1} \ldots j_{N}} \prod_{m} p_{j_{m}} |\langle j_{m}|\theta_{n_{m}}\rangle |^2 \\
& = & \bigg(\sum_{\{j_{m}\}_{m\neq l}} \prod_{m\neq l} p_{j_{m}} |\langle j_{m}|\theta_{n_{m}} \rangle |^2  \bigg) \\
&& \qquad \times \; \bigg( \sum_{j_{l}} p_{j_{l}} |\langle j_{l} | \theta_{n_{l}} \rangle|^{2} \bigg).
\end{eqnarray*}
So the eigenvalues of ${\cal L}(\rho')$ for the $N$ qubit probe undergoing non-entangling evolution is
\begin{eqnarray}
\label{eq:lambdaN}
\frac{1}{\lambda_{\vec{n}}}&=& \sum_{l=1}^{N} \frac{\sum_{j_{l} k _{l}}p_{j_{l}}\frac{p_{j_{l}}-p_{k_{l}}}{p_{j_{l}}+p_{k_{l}}}  \langle j_{l}|\sigma_3 |k_{l}\rangle \langle j_{l}|\theta_{n_{l}}\rangle \langle \theta_{n_{l}}|k_{l}\rangle   }{ \sum_{j_{l}} p_{j_{l}} |\langle j_{l} | \theta_{n_{l}} \rangle |^{2}} \nonumber\\
&=& \sum_{l=1}^{N} \frac{1}{\lambda_{n_{l}}}.
\end{eqnarray}

\subsection{\texorpdfstring{$\cal L$}{L} for non-entangling dynamics \label{app1b}}

Using the expression for $1/\lambda_{\vec{n}}$ from Eq.~(\ref{eq:lambdaN}) and the fact that corresponding to the optimal state of the one qubit probe, the eigenvalues of ${\cal L}$ are $\mp 1$, we can write the symmetric logarithmic derivative operator on the $N$ qubit probe as
\begin{equation}
\label{eq:L2a}
{\cal L} =-N|+\cdots + \rangle\langle +\cdots+| 
+\cdots+ N |- \cdots - \rangle\langle - \cdots -|
\end{equation}
 Using $ |+\rangle\langle+|=(\openone+\sigma_1)/2$ and $|-\rangle\langle-|=(\openone-\sigma_1)/2$ we get,
\[ {\cal L} = -\frac{1}{2^N}\sum_{r=0}^{N} (N-2r) \big[ \hat{\cal C}(\openone+\sigma_1)^{\otimes (N-r)}\otimes(\openone-\sigma_1)^{\otimes r} \big], \]
 where $r$ is the number of $|-\rangle \langle -|$ projectors in each term in Eq.~(\ref{eq:L2a}) and $\hat{\cal C}$ is a shorthand indicating all terms that are tensor products of $N-r$ factors of $(\openone + \sigma_{1})$ and $r$ factors of $(\openone - \sigma_{1})$.
  
Once the sum is distributed over the tensor product, we get terms with $N-q$ factors that are  $\openone$'s and $q$ $\sigma_1$'s with $q=0,1,\ldots,N$. We first focus on the sign of the various terms with fixed number of $\openone$'s and $\sigma_{1}$'s in the expression for ${\cal L}$. If out of the  $q$ factors of $\sigma_1$'s, an odd number $s$ of them come from the  $r$  $(\openone -\sigma_1)$ factors then the term, as a whole is negative.  Now, out of the total $q$ factors of $\sigma_{1}$ we can pick $s$ of them in $^{q}C_{s}$ ways. Now, out of the $N-q$ factors of $\openone$, $r-s$ of them have to come from the $(\openone - \sigma_{1})$ terms. These $r-s$ factors can be picked in ${}^{N-q}C_{r-s}$ ways. The remaining $\openone$'s and $\sigma_{1}$'s come from the $(\openone + \sigma_{1})$ terms. So the total number of ways in which one can construct a term with $q$  $\sigma_1$'s starting from a term with $r$ factors of $(\openone -\sigma_1)$ is 
\[ {}^N\! C_r-2\sum_{s=1,3,\cdots}^{{\rm min}(r,q)} {}^{N-q} C_{r-s} {}^q  C_s. \]
We have for all $q\geq 0$ 
\[ -\sum_{r=0}^{N} (N-2r) {}^N \!C_r =  -N 2^N+2N\sum_{r=0}^{N-1}\;^{N-1}C_r=0 \]
Hence
\begin{eqnarray}
{\cal L}&=& \frac{2}{2^N}\sum_{q=1}^{N}\sum_{r=0}^{N}\sum_{s=1,3,\cdots}^{{\rm min}(r,q)} (N-2r) \;^{N-q} C_{r-s} \;^q C_s\nonumber\\
& & \times \{\hat{\cal C} \openone^{\otimes (N-q)}\otimes \sigma_1 ^{\otimes q} \}\nonumber
\end{eqnarray} 
For $q=1$ the summation in the $\cal L$ reduces to 
\[  2\sum_{r=1}^{N} (N-2r)\;^{N-1}C_{r-1}=-2^N \]
Now for any $q>1$, we have
\begin{eqnarray}
 \sum_{r=1}^{N}   \sum^{\min (q,r)}_{s=1,3,\cdots} & & (N-2r) {}^{N-q} C_{r-s} {}^q C_s  \qquad \nonumber \\
 & & =  N\sum_{r=1}^{N} \sum_{s=1,3,\cdots}^{\min(q,r)} {}^{N-q} C_{r-s}  {}^qC_s\nonumber\\
& & \qquad -2(N-q)\sum_{r=1}^{N} \sum_{s=1,3,\cdots}^{\min(q,r)} \!\! {}^{N-q-1} C_{r-s-1} {}^{q}C_s\nonumber\\ 
& &  \qquad -2q\sum_{r=1}^{N} \sum_{s=1,3,\cdots}^{\min (q,r)} {}^{N-q} C_{r-s} \;^{q-1} C_{s-1}
\label{eqnsum}
\end{eqnarray}
To compute the sums in the equation above, we use the following results: For $q>1$, $\sum_{i=1,3,\cdots}^q {}^qC_i = 2^{q-1}$  and $\sum_{i=0,2,\cdots}^{q-1} {}^{q-1}C_{i}=2^{q-2}$. Eq.~(\ref{eqnsum}) becomes
\begin{eqnarray*}
\sum_{r=1}^{N} \sum_{s=1,3,\cdots}^{\min(q,r)} & & (N-2r) \;^{N-q} C_{r-s} \;^q C_s \\
& &=N\sum_{i=1,3,\cdots}^q {}^qC_i \sum_{j=0,1,\cdots}^{N-q}\;^{N-q}C_{j} \\
& &\;-2(N-q)\sum_{i=1,3,\cdots}^q {}^qC_i \sum_{j=0,1,\cdots}^{N-q} {}^{N-q}C_{j} \\
& &\;-2q\sum_{i=0,2,\cdots}^{q-1} {}^{q-1}C_{i}\sum_{j=0,1,\cdots}^{N-q-1} {}^{N-q-1}C_{j}=0 
\end{eqnarray*}
Thus the symmetric logrithmic derivative for an $N$ qubit probe evolving under a non entangling Hamiltonian is
\[ {\cal L} = -\{\hat{\cal C} \openone^{\otimes (N-1)}\otimes \sigma_1\}. \]

\section{Entangling Hamiltonian \label{app2}}

When the parameter dependent evolution is generated by the entangling Hamiltonian, $\sigma_{3}^{\otimes N}/2$ and we consider an initial state of the probe of the form $\tilde{\rho}_{X}^{\otimes N}$, we have
\[ \rho' =  -\frac{i}{2}  \big[ (\sigma_3\rho)^{\otimes N}-(\rho\sigma_3)^{\otimes N} \big] \]
Now
\begin{eqnarray}
\rho_{\vec{j},\vec{k}}'&=&-\frac{i}{2}\prod_{l}\langle j_{l}|\sigma_3\rho|k_{l}\rangle +\frac{i}{2}\prod_{l} \langle j_{l}|\rho\sigma_3|k_{l}\rangle \nonumber\\
&=&-\frac{i}{2}\prod_{l}p_{j_{l}}\langle j_l|\sigma_3|k_{l}\rangle +\frac{i}{2}\prod_{l}p_{j_{l}} \langle j_l|\sigma_3|k_{l}\rangle \nonumber
\end{eqnarray}
using the above, we obtain
\begin{eqnarray}
{\cal L}(\rho')&=&-i \sum_{\{j_l,k_l\}} \frac{\prod_{l} p_{k_l}-\prod_{l}p_{j_l}}{\prod_l p_{k_l}+\prod_l p_{j_l}} \prod_{l}\langle j_l|\sigma_3|k_l\rangle\nonumber\\
& & \times |j_{1}, j_{2}, \ldots j_{N} \rangle \langle k_{1}, k_{2}, \ldots k_{N}|\nonumber,
\end{eqnarray}
and
\begin{eqnarray}
{\rm tr}(\rho E_{\vec{n}} {\cal L}(\rho'))&=&-i \sum_{\{j_l,k_l\}} \frac{\prod_{l} p_{k_l}(\prod_{i} p_{k_l} -\prod_{i}p_{j_l})}{\prod_l p_{k_l}+\prod_l p_{j_l}} \nonumber\\
& & \prod_{l} \langle j_l|\sigma_3| k_l\rangle \langle k_l|\theta_{n_{l}}\rangle  \langle \theta_{n_{l}}|j_l\rangle.
\label{eq:B1a}
\end{eqnarray}
We also have
\begin{equation}
\label{eq:B2a}
{\rm tr}(E_{\vec{n}}\rho)=\sum_{\{j_{l}\}}\prod_{l} p_{j_l} |\langle j_{l}|\theta_{n_{l}} \rangle|^2 
\end{equation}
We get $1/\lambda_{\vec{n}}$ by dividing the right hand side of Eq.~(\ref{eq:B1a}) by that of Eq.~(\ref{eq:B2a}). Note that the denominator in (\ref{eq:B2a}) is real and so is the first part of each term in the double sum in (\ref{eq:B1a}). So the product term $\prod_{l}\langle j_l|\sigma_3| k_l\rangle \langle k_l|\theta_{n_{l}}\rangle  \langle \theta_{n_{l}}|j_l\rangle$, has to be pure imaginary for $1/\lambda_{\vec{n}}$ to be real as required for saturating the bound on the Fisher information as discussed in Section.~\ref{sec2}. However each term in this product comes form each qubit in the probe. So if we assume that each qubit in the probe is in the optimal state $\tilde{\rho}_{X}$ corresponding to the $N=1$ state, then for each qubit $\langle j|\sigma_3| k\rangle \langle k|\theta_n\rangle  \langle \theta_n|j\rangle$ has to be pure imaginary so that again, the bound is saturated as assumed. This implies that when $N$ is even then $\lambda_{\vec{n}}$ are all pure imaginary and so the tensor product state $\tilde{\rho}_{X}^{\otimes N}$ is not the optimal state of the probe corresponding to the entangling dynamics and readout procedure that we are considering when $N$ is even. 

\subsection{Optimal state of a two qubit probe \label{app2a}}
 
 Using the same notation as in Appendix~\ref{app1}, Eq.~(\ref{eq:sol1}) reduces to the following sixteen algebraic equations, 
\begin{eqnarray}
	\label{eq:2Qnent2}
	K_{+++}+K_{+--}a_{1}+ K_{-+-}b_{1}  + K_{--+}c_{11}&=&  0 \nonumber \\
	K_{--+}+ K_{-+-}a_{1}+ K_{+--}b_{1}  +K_{+++}c_{11}&=& 0\nonumber \\
         	K_{+--} + K_{+++} a_{1} + K_{--+} b_{1} + K_{-+-} c_{11} & = & -4c_{23} \nonumber \\
	K_{-+-} + K_{--+}a_{1} + K_{+++} b_{1} + K_{+--} c_{11} & = & - 4c_{32}  \nonumber \\
	K_{+++} a_{3} + K_{-+-} c_{31} & = & 0 \nonumber \\
	K_{+++} c_{22} - K_{--+} c_{33} & = & 0 \nonumber \\
	K_{+++} b_{2} + K_{+--}c_{12} & = &  4c_{31}  \nonumber \\
	K_{+++} a_{2} + K_{-+-} c_{21} & = & 4 c_{13} \nonumber \\
	K_{-+-} a_{3} + K_{+++} c_{31} & = & -4 b_{2} \nonumber \\
	K_{+++} b_{3} + K_{+--} c_{13} & = & 0 \nonumber \\
	K_{+--} b_{3} + K_{+++} c_{13} & = & -4a_{2} \nonumber \\
	K_{+++} c_{33} - K_{--+} c_{22} & = & 0 \nonumber \\
	K_{+++} c_{21} +K_{-+-} a_{2}& = & 0 \nonumber \\
	K_{+++} c_{12}+K_{+--} b_{2} & = & 0\nonumber \\
	K_{+++} c_{32}+K_{--+} c_{23} & = & 4 b_{1} \nonumber \\
	K_{+++} c_{23} + K_{--+} c_{32}& = &4 a_{1} .
	\end{eqnarray}
In this case, as noted earlier it is easy to verify that $ \tilde{\rho}^{(2)}_{X}  =  \tilde{\rho}_{X} \otimes \tilde{\rho}_{X} $ does not give any solution for $1/\lambda_{n}$. Solving these equations for the variables $a_i,b_j \ and \ c_{ij}$ and applying in the general form of $\rho$ for 2 qubits we get the optimal initial state for the two qubit probe. Multiple solutions are allowed and for maximizing $\langle {\cal L}^{2} \rangle$ we look for pure state solutions. One pure state solution may be obtained by setting all $a_{i}$, $b_{j}$ and $c_{ij}$ to zero except for $c_{11} =1$ and $c_{23} = c_{32} = \pm 1$ which then leads to
\[\frac{1}{\lambda_{++}}= -\frac{1}{\lambda_{--}} = -1 \quad {\rm and } \quad  \frac{1}{\lambda_{+-}}= \frac{1}{\lambda_{-+}} =c,\]
where $c$ can be any real number, including 0. 

\section{\texorpdfstring{$1/\lambda$}{Lambda} for odd \texorpdfstring{$N$}{N} \label{app3} }

Starting from Eq.~(\ref{eq:B1a}) by that of Eq.~(\ref{eq:B2a}) we find the eigenvalues of $\cal L (\rho')$ for an $N$ qubit probe undergoing entangling evolution as, 
\begin{eqnarray} 
\frac{1}{\lambda_{\vec{n}}} &=&\frac{-i \sum_{\{j_l,k_l\}} \frac{\prod_{l} p_{k_l}(\prod_{i} p_{k_l} -\prod_{i}p_{j_l})}{\prod_l p_{k_l}+\prod_l p_{j_l}} }{\sum_{\{j_{l}\}}\prod_{l} p_{j_l} |\langle j_{l}|\theta_{n_{l}} \rangle|^2 } \nonumber\\
& & \times \prod_{l} \langle j_l|\sigma_3| k_l\rangle \langle k_l|\theta_{n_{l}}\rangle  \langle \theta_{n_{l}}|j_l\rangle.
\label{y4}
\end{eqnarray}
The optimal single qubit state $\tilde{\rho}_{X}$ is diagonal in the eigenbasis $\{|i\rangle, \, |\bar{i}\rangle \}$ of the $\sigma_{2}$ operator and so we can write it as,
\[ \tilde{\rho}_{X}=p_i|i\rangle\langle i|+p_{\bar{i}}|\bar{i}\rangle\langle \bar{i}|, \]
so that $|j_{l}\rangle, |k_{l}\rangle = \{|i\rangle ,\, |\bar{i}\rangle\}$. As before the readout operators, $|\theta_{n_{l}}\rangle \langle \theta_{n_{l}}|\;'$s are given by $|+\rangle \langle +|$ and $|-\rangle \langle -|$. Using the inner products, $\langle i|\pm\rangle=(1\mp i)/2$ , $\langle \bar{i}|\pm\rangle=(1\pm i)/2$, $\langle i|+\rangle \langle +|\bar{i}\rangle=-i/2$, and $\langle i|-\rangle \langle -|\bar{i}\rangle=i/2$ we get for any choice of $\theta_n\;'$s,
\begin{equation}
\label{eq:y5}
\sum_{\{j_{l}\}}\prod_{l} p_{j_l} |\langle j_{l}|\theta_{n_{l}} \rangle|^2 = \frac{1}{2^N} \left(p_i+p_{\bar{i}}\right)^N 
\end{equation}

Since $\langle i |\sigma_3|i\rangle=\langle \bar{i} |\sigma_3|\bar{i}\rangle=0$ and $\langle i|\sigma_3|\bar{i}\rangle=1$, only terms with $\vec{j} = \vec{\bar{k}}$ contribute to the expression for $\rho'_{\vec{j} \vec{k}}$ and hence to ${\cal L}(\rho')$ and $1/\lambda_{n}$ as well. Consider a single term in the double sum over ${\{ j_{l}, \,k_{l}\}}$ in the numerator of equation (\ref{y4}), where
 \[ |\vec{j}\rangle = |i,i,\cdots,i,\bar{i}\rangle \quad {\rm and} \quad  |\vec{k}\rangle=|\bar{i},\bar{i},\cdots ,\bar{i},i\rangle.\] 
 In this term we have the factor
 \begin{eqnarray}
 \prod_{l} \langle j_{l}|\theta_{n_{l}}\rangle  \langle \theta_{n_{l}}|k_{l}\rangle &=&{\mp\frac{i}{2} \cdot \mp\frac{i}{2}\cdots\mp\frac{i}{2} \cdot \pm\frac{i}{2}}\nonumber\\
 &=& \frac{i^N}{2^N}(\mp \cdot \mp\cdots\mp \cdot \pm)\nonumber  
 \end{eqnarray}
 In the double sum over ${\{j_{l},\, k_{l}\}}$ in equation (\ref{y4}), there will be ${}^N\!C_1$ terms each having $N-1$ $(\mp)$ terms and one $(\pm)$ term. In general, there will be ${}^N\! C_r$ terms each having $N-r$ $(\mp)$ terms and $r$ $(\pm)$ terms. When $E_{\vec{n}}$ and hence $\theta_{n_{l}}$ is fixed, then every term with $N-r$ $\mp$ terms and $r$ $\pm$ terms has the same sign and so they can all be grouped together. We represent these terms that are grouped together as  $\{\underbrace{\mp\cdots\mp}_{N-r} \underbrace{\pm\cdots\pm}_{r}\}$, since the group is labeled by $r$. Using this notation and Eq.~(\ref{eq:y5}), we have
\begin{eqnarray}
\frac{1}{\lambda_{\vec{n}}} &=&Q\frac{p_N^0-p_0^N}{p_N^0+p_0^N}\;^NC_0 p_N^0\{\mp\mp\cdots\mp\}\nonumber\\
& & +Q\frac{p_{N-1}^{1}-p_{1}^{N-1}}{p_{N-1}^{1}+p_{1}^{N-1}} {}^N \! C_1 p_{N-1}^1\{\mp\mp\cdots\mp\pm\}\nonumber\\
& & +\cdots\nonumber\\
& & +Q\frac{p_{N-r}^{r}-p_{r}^{N-r}}{p_{N-r}^{r}+p_{r}^{N-r}} {}^{N} \! C_{r}p_{N-r}^{r}\{\mp\cdots\mp\pm\cdots\pm\}\nonumber\\
& & +\cdots\nonumber\\
& & -Q\frac{p_{N-r}^{r}-p_{r}^{N-r}}{p_{N-r}^{r}+p_{r}^{N-r}} {}^{N} \! C_{N-r}p_{r}^{N-r} \{\pm\cdots\pm\mp\cdots\mp\}\nonumber\\
& & -\cdots\nonumber\\
& & -Q\frac{p_{N-1}^{1}-p_{1}^{N-1}}{p_{N-1}^{1}+p_{1}^{N-1}} {}^N \! C_{N-1}p_1^{N-1} \{\pm\pm\cdots\pm\mp\}\nonumber\\
& & -Q\frac{p_N^0-p_0^N}{p_N^0+p_0^N} {}^N \! C_{N}p_0^N \{\pm\pm\cdots\pm\}\nonumber
\end{eqnarray}
 where $Q=\frac{-i^{N+1}}{\left(p_i+p_{\bar{i}}\right)^N}$ and $p_{N-r}^{r}=(p_{i})^{N-r}\;(p_{\bar{i}})^{r}$. Now consider the first and last terms of the sum as a pair:
 \begin{eqnarray}\label{y5}
A_1 &=& Q\frac{p_N^0-p_0^N}{p_N^0+p_0^N} {}^N \! C_0 p_N^0\{\mp\mp\cdots\mp\}\nonumber\\
& & -Q\frac{p_N^0-p_0^N}{p_N^0+p_0^N} {}^N \!C_{N}p_0^N \{\pm\pm\cdots\pm\} 
\end{eqnarray}
Since $N$ is odd, for a given choice of $E_{\vec{n}}$, $\{\mp\mp\cdots\mp\}$ and $\{\pm\pm\cdots\pm\}$ have opposite signs. Hence, equation (\ref{y5}) reduces to
\begin{eqnarray}
A_1 &=& \pm Q \left({}^N \! C_0 p_N^0-{}^N\!C_N p_0^N\right) \nonumber 
\end{eqnarray}
Simillarly, we can pair up the remaining terms. Since for even and odd $r$ 
\[ Q\frac{p_{N-r}^{r}-p_{r}^{N-r}}{p_{N-r}^{r}+p_{r}^{N-r}} {}^{N}\!C_{r}p_{N-r}^{r}\{\mp\cdots\mp\pm\cdots\pm\}$$ for any $E_{\vec{n}}$'s has opposite signs, we get
\begin{eqnarray}
\frac{1}{\lambda_{\vec{n}}} &=& \pm \frac{i^{N+3}\left(p_i-p_{\bar{i}}\right)^N}{\left(p_i+p_{\bar{i}}\right)^N}  \nonumber 
\end{eqnarray}
The sign of $\lambda_{\vec{n}}$ depends on the choice of $E_{\vec{n}}$s. Using $1/\lambda_{n_{l}} = \pm(p_{i} - p_{\bar{i}})/(p_{i} + p_{\bar{i}})$, we see that for any odd $N$, 
\[ \frac{1}{\lambda_{\vec{n}}} = \frac{i^{N+3}}{\lambda_{n1} \lambda_{n_{2}} \cdots \lambda_{n_{N}}}. \]

Generalizing the above result, the best initial state for any $2^n(2m+1)$ number of qubits can be obtained as $\rho=\rho_{2^n}^{\otimes(2m+1)}$, where $m,n=0,1,2,\cdots$ and $\rho_{2^n}$ is the best initial state of $2^n$ qubits.

\bibliography{inverseCR}

\end{document}